\documentclass[12pt]{article}

\usepackage{cite}
\usepackage{hyperref}
\usepackage{epsfig}
\usepackage{graphics}
\usepackage{inputenc}
\usepackage{color}
\inputencoding{latin1}

\def\to{\ensuremath{\rightarrow}}

\newcommand{\ariadne}{A\scalebox{0.8}{RIADNE}}
\newcommand{\herwig}{\scalebox{0.8}{HERWIG}}
\newcommand{\pythia}{P\scalebox{0.8}{YTHIA}}
\newcommand{\diclus}{\scalebox{0.8}{DICLUS}}
\newcommand{\as}{\ensuremath{\alpha_{\mathrm{s}}}}
\newcommand{\asi}[1]{\ensuremath{\alpha_{\mathrm{s}#1}}}

\newcommand{\kT}{\ensuremath{k_{\perp}}}
\newcommand{\pT}{\ensuremath{p_{\perp}}}

\newcommand{\pTi}[1]{\ensuremath{p_{\perp #1}}}

\newcommand{\particle}[1]{\ensuremath{\mathrm{#1}}}
\newcommand{\antiparticle}[1]{\ensuremath{\bar{\mathrm{#1}}}}
\newcommand{\g}{\particle{g}}
\newcommand{\p}{\particle{p}}
\newcommand{\q}{\particle{q}}

\newcommand{\W}{\particle{W}}

\newcommand{\qbar}{\antiparticle{q}}

\newcommand{\e}{\ensuremath{\mathrm{e}^-}}
\newcommand{\ebar}{\ensuremath{\mathrm{e}^+}}
\def\ee{\ebar\e}

\newcommand{\lam}{\ensuremath{\Lambda_{\mathrm{QCD}}}}

\def\sub#1{\ensuremath{_{\mathrm{#1}}}}
\def\sup#1{\ensuremath{^{\mathrm{#1}}}}
\def\ordo#1{\ensuremath{{\cal O}(#1)}}
\def\sud#1{\ensuremath{\Delta_{S_{#1}}}}

\begin{document}

\sloppy

\begin{titlepage}

  \begin{flushright}
    LU TP 01--38\\
    hep-ph/0112284 \\
    December 2001\\
    Revised May 2002
  \end{flushright}
  \begin{center}
    
    \vskip 10mm {\Large\bf\boldmath Correcting the Colour-Dipole
      Cascade Model with Fixed Order Matrix Elements} \vskip 15mm

    {\large Leif Lönnblad}\\
    Dept.~of Theoretical Physics,\\
    S\"olvegatan 14A, S-223 62  Lund, Sweden\\
    Leif.Lonnblad@thep.lu.se

  \end{center}
  \vfill
  \begin{abstract}
    
    An algorithm is presented in which the Colour-Dipole Cascade Model
    as implemented in the \ariadne\ program is corrected to match the
    fixed order tree-level matrix elements for $\ee\rightarrow n$
    jets. The result is a full parton level generator for \ee\ 
    annihilation where the generated states are correct on tree-level
    to fixed order in \as\ \emph{and} to all orders with modified
    leading logarithmic (MLLA) accuracy. In this paper, matrix
    elements are used up to second order in \as, but the scheme is
    applicable also for higher orders.  A strategy for also including
    exact virtual corrections to fixed order is suggested and the
    possibility to extend the scheme to hadronic collisions is
    discussed.

  \end{abstract}

\end{titlepage}

\section{Introduction}
\label{sec:intro}

Perturbative QCD has been very successful in describing many features
of multi-particle production in high energy collisions. There are,
however, several problems which have not yet been solved, mostly
related to the transition between the perturbative and
non-perturbative description of the theory. Observables involving a
few widely separated jets are in principle well described with
fixed-order perturbative matrix elements (MEs) for producing a few
partons.  But to make precision comparisons with experiments, it is
important to understand the transition of these partons to observable
hadrons.  Our best knowledge of this transition comes from
hadronization models which describes how multi-parton states are
transformed into multi-hadron ones. But for these models to work
reliably one needs also a description of the collinear and soft
partons describing the internal structure of widely separated jets and
the soft partons between the jets.

To describe soft and collinear partons it is not feasible to use
fixed-order perturbation theory. Not only do the MEs for many-parton
states become extremely complicated but, since the partons are no
longer widely separated, the increase in phase space introduces large
logarithms which compensates the smallness of \as\ and makes the whole
perturbative expansion ill-behaved. To describe the inner structure of
jets, a more practical approach is to use a parton shower (PS)
procedure.  Here the large logarithms are resummed to all orders at
the expense of only keeping the leading logarithmic behaviour of the
full matrix elements.

To get a near complete description of multi-particle production it
would be desirable to combine the generation of a few widely separated
partons according to fixed-order MEs with the evolution of these
states according PSs and finally the transition into hadrons using a
hadronization model. To do this is, however, highly non-trivial and so
far there exist no general procedure which is entirely satisfactory.
The main problem is that one needs a resolution scale to separate the
ME generation from the PS one. This scale needs to be small enough to
benefit from the full ME description, but if it becomes too small the
final result is spoiled by non-physical large logarithms involving the
separation scale.

For some special cases such as e.g.\ $\ee\rightarrow3$~jets
\cite{Gustafson:1988rq,Lonnblad:1992tz,Bengtsson:1987hr,
  Bengtsson:1987et,Seymour:1995we,Seymour:1995df},
$\e\p\rightarrow2+1$~jets\cite{Seymour:1994ti,Lonnblad:1995wk} and
$\p\p\rightarrow\W+1$~jet\cite{Lonnblad:1996ex,Miu:1998ju,Mrenna:1999mq,
  Corcella:1999gs}, there are working procedures for combining ME and
PS.  Recently there has also been suggested a couple of more general
procedures
\cite{Andre:1998vh,Norrbin:2000uu,Potter:2000an,Catani:2001cc,
  Collins:2000qd,Chen:2001ci}, but none of them can be considered to
be the final word on the matter.  In this paper the procedure by
Catani and collaborators in \cite{Catani:2001cc} is taken as a
starting point to create a matching between fixed order ME generators
and the colour-dipole cascade model (CDM)
\cite{Gustafson:1986db,Gustafson:1988rq} as implemented in the
\ariadne\cite{Lonnblad:1992tz} event generator.  The resulting
algorithm is also not complete, but it provides a cleaner interface
between the ME and PS stages of the generation, and it carefully
treats the influence of the available phase space on the Sudakov form
factors needed to correct the ME generation.

The outline of this paper is as follows. In section
\ref{sec:matr-elem-part} the algorithm of Catani et al.\ is presented,
reformulated to better fit the CDM. Then, in section
\ref{sec:reconstr-emiss}, the reconstruction of the resolution scales
of a partonic state generated with MEs is discussed, followed by the
description in section \ref{sec:sudak-veto-algor} of how the Sudakov
form factors are calculated and a step by step description of the
whole algorithm in section \ref{sec:procedure-step-step}. In section
\ref{sec:results} a few initial results from the new algorithm for
$\ee\rightarrow n$~jets with $n\le4$ are presented. Finally, in
section \ref{sec:conclusions-outlook}, some conclusions are presented
together with a discussion on how the procedure can be extended to
also be used for collisions involving hadrons in the initial state.
Also the possibility to improve the description to get virtual
corrections exact to fixed-order perturbation theory is discussed.

\section{Matrix elements and parton cascades}
\label{sec:matr-elem-part}

To describe \ee\ annihilation into $n$ jets with fixed order
perturbation theory one needs to introduce a resolution scale to avoid
divergencies in the ME. This scale can be defined in many different
ways, usually connected to a specific jet reconstruction algorithm
which can be used both on the ME and on the final-state hadrons
observed in an experiment. Given such a resolution scale $Q_0$ and a
maximum scale, $Q$, set by the total center of mass energy, we can
write down the fraction of $n$ jet events for $n\le4$, $R_n(Q,Q_0)$,
given by the second order MEs as
\begin{eqnarray}
  \label{eq:RnME}
  R_2(Q,Q_0) & = &
  1 + \as C_{0,1}(Q, Q_0) + \as^2C_{0,2}(Q, Q_0),\nonumber\\
  R_3(Q,Q_0) & = &
  \as C_{1,1}(Q,Q_0)( 1 + \as C_{1,2}(Q, Q_0) ),\\
  R_4(Q,Q_0) & = & \as^2 C_{2,2}(Q,Q_0),\nonumber
\end{eqnarray}
where $R_4$ receives contributions from \q\qbar\g\g\ as well as from
$\q\qbar\q'\qbar'$ final states. The coefficients $C_{n,m}(Q, Q_0)$
are related to the emission of $n$ partons to $\ordo{\as^m$}, i.e.\ 
$C_{n,n}$ corresponds to resolved tree-level diagrams and $C_{n,m}$,
with $m>n$, corresponds to the sum of virtual diagrams and unresolved
emission diagrams. The problem with a small resolution scale comes
about because the coefficients contains logarithms of $Q/Q_0$ which,
for small $Q_0$, destroys the \as-expansion. In fact $R_3$ above
becomes negative for small enough $Q_0$.

The parton shower approach is quite different. Here gluons are emitted
or split into $\q\qbar$ pairs in according to simple $1\to2$ (or, as
in the CDM case, $2\to3$) splitting functions in an ordered
probabilistic cascade.  The emissions are hence ordered according to
an associated evolution scale so that the first emissions has the
highest scale and subsequent emissions decreasingly lower scales. The
definition of the evolution scale, $\rho$, is different in different
shower models, e.g.\ invariant mass in \pythia\cite{Sjostrand:2000wi}
and invariant transverse momentum in \ariadne, but common for all
models is the occurrence of a Sudakov form factor corresponding to the
probability of having no emissions between two given scales. If
$\as{\cal P}(\rho,z,\phi)$ is the splitting probability for emitting a
gluon from the initial $\q\qbar$ pair, the two-jet rate above some
scale $\rho_0$ becomes
\begin{equation}
  \label{eq:R2PSSud}
  R_2(\rho_{m},\rho_0) = \sud{2}(\rho_{m},\rho_0) =
  \exp\left(-\as\int^{\rho_{m}}_{\rho_0}d\rho \int dz\int d\phi
    {\cal P}(\rho,z,\phi)\right),
\end{equation}
where $\rho_m$ is the maximum kinematically allowed scale, and the $z$
and $\phi$ integrals are over all kinematically allowed values.
Performing the integrals and expanding the exponential, the two-jet
rate can be expressed in the same form as the matrix element in
eq.~(\ref{eq:RnME})
\begin{equation}
  \label{eq:R2PS}
  R_2(\rho_m,\rho_0) =
  1 + \as C\sup{PS}_{0,1}(\rho_, \rho_0) +
  \as^2C\sup{PS}_{0,2}(\rho_m, \rho_0) + \ldots
\end{equation}
Similarly the three-jet rate according to a parton shower
scenario can be written as
\begin{equation}
  \label{eq:eq:R3PSSud}
  R_3(\rho_m,\rho_0) = \as\int^{\rho_m}_{\rho_0}d\rho \int dz\int d\phi
    {\cal P}(\rho,z,\phi)\sud{2}(\rho_{m},\rho)\sud{3}(\rho,\rho_0),
\end{equation}
i.e.\ the probability of emitting a gluon at some scale $\rho$, times
the probability that no gluon was emitted above (before) $\rho$ and
the probability that nothing was emitted below. The latter is also a
Sudakov form factor, symbolically written $\sud{3}(\rho,\rho_0)$,
giving the probability of not having any further emissions from the
\q\g\qbar\ state above the scale $\rho_0$ and will in general depends on
the full kinematics of the three-parton state. Still, it can be written
as an exponential of minus an emission probability integrated over the
allowed phase space. It is therefore easily expanded in powers of \as\
and, after integrating an expression similar to the one in
eq.~(\ref{eq:RnME}) can be obtained:
\begin{equation}
  \label{eq:R3PS}
  R_3(\rho_m,\rho_0) =
  \as C\sup{PS}_{1,1}(\rho_, \rho_0) ( 1 + 
  \as C\sup{PS}_{1,2}(\rho_m, \rho_0) + \ldots ).
\end{equation}

Note that, as in eq.~(\ref{eq:RnME}), the coefficient $C\sup{PS}_{1,2}$
can become large and negative, but since it is the result of an
expansion of an exponential which is always positive and below one,
the $R_3$ will always be positive and finite.

For the four-jet rate there is an additional complication since there
are both \q\qbar\g\g- and \q\qbar\q'\qbar'-states, but it can be
written on the form
\begin{eqnarray}
  \label{eq:R4PSSud}
  R_4(\rho_m,\rho_0)  &=&
  \as^2\int^{\rho_m}_{\rho_0}d\rho_1\int dz_1\int d\phi_1
  {\cal P}(\rho_1,z_1,\phi_1)\sud{2}(\rho_{m},\rho_1)
  \sud{3}(\rho_1,\rho_2)\times\nonumber\\
  & & \rule{2cm}{0mm} \int_{\rho_0}^{\rho_1}d\rho_2\int dz_2 \int d\phi_2
  \left[{\cal P}_{\g\g}(\rho_2,z_2,\phi_2)
    \sud{\q\g\g\qbar}(\rho_2,\rho_0)\right.\\
  & & \rule{5.26cm}{0mm} +\left.{\cal P}_{\q\qbar}(\rho_2,z_2,\phi_2)
    \sud{\q\qbar\q'\qbar'}(\rho_2,\rho_0)\right], \nonumber
\end{eqnarray}
where ${\cal P}_{\g\g}$ is the splitting function for emitting an
additional gluon from the \q\g\qbar\ state and ${\cal P}_{\q\qbar}$ is
the splitting function for splitting the gluon in the \q\g\qbar\ state
into a new quark-antiquark pair.

Again, the Sudakovs, $\sud{\q\g\g\qbar}$ and $\sud{\q\qbar\q'\qbar'}$,
gives the no-emission probability from the \q\g\g\qbar\ and
\q\qbar\q'\qbar' states respectively and can be written as simple
exponentials, which can be expanded in \as\ and, after integration, an
expression similar to the one in eq.~(\ref{eq:RnME}) can be written
\begin{equation}
  \label{eq:R4PS}
  R_4(\rho_m,\rho_0) =
  \as^2 C\sup{PS}_{2,2}(\rho_, \rho_0) ( 1 + 
  \as C\sup{PS}_{2,3}(\rho_m, \rho_0) + \ldots ).
\end{equation}

The coefficients $C\sup{PS}_{n,m}$ contains only the leading
logarithmic parts\footnote{In fact, most parton shower implementations
  also contain some non-leading terms corresponding to the
  \emph{modified} leading logarithmic approximation (MLLA)
  \cite{Mueller:1983cq,Mueller:1984cq,Dokshitzer:1984dx}. In
  particular, for the case of the \ariadne\ program,
  $C\sup{PS}_{1,1}=C_{1,1}$ and $C\sup{PS}_{0,1}=C_{0,1}$} of the
corresponding exact ones, $C_{n,m}$. To arrive at a combined parton
shower and matrix element generator would then mean using the parton
shower method but replacing the first few coefficients
$C\sup{PS}_{n,m}$ with the exact ones. Here, a more modest approach
will be taken, where only the coefficients $C\sup{PS}_{n,n}$,
corresponding to the resolved tree-level diagrams will be replaced by
the exact versions. This amounts to generating events according to the
$n$-jet rates
\begin{eqnarray}
  \label{eq:RnMEPS}
  R_2(\rho_m,\rho_0) &=&\sud{2}(\rho_{m},\rho_0),\nonumber\\
  R_3(\rho_m,\rho_0) &=& \as\int^{\rho_m}_{\rho_0}d\rho\,\,
  c\sup{ME}_{1,1}(\rho_m,\rho)\sud{2}(\rho_{m},\rho)\sud{3}(\rho,\rho_0),\\
  R_{\q\g\g\qbar}(\rho_m,\rho_0) &=&
  \as^2\int^{\rho_m}_{\rho_0}d\rho_1\int^{\rho_1}_{\rho_0}d\rho_2\,\,
  c\sup{ME}_{2,2(\q\g\g\qbar)}(\rho_m,\rho_1,\rho_2)\times\nonumber\\
  & & \rule{4cm}{0mm} \sud{2}(\rho_{m},\rho_1)
  \sud{3}(\rho_1,\rho_2)\sud{\q\g\g\qbar}(\rho_2,\rho_0),\nonumber\\
  R_{\q\qbar\q'\qbar'}(\rho_m,\rho_0) &=&
  \as^2\int^{\rho_m}_{\rho_0}d\rho_1\int^{\rho_1}_{\rho_0}d\rho_2\,\,
  c\sup{ME}_{2,2(\q\qbar\q'\qbar')}(\rho_m,\rho_1,\rho_2)\times\nonumber\\
  & & \rule{4cm}{0mm} \sud{2}(\rho_{m},\rho_1)
  \sud{3}(\rho_1,\rho_2)\sud{\q\qbar\q'\qbar'}(\rho_2,\rho_0),\nonumber
\end{eqnarray}
where the $c_{n,n}\sup{ME}$ are the differential full tree-level
matrix elements (the $z$ and $\phi$ integrals have been left out for
brevity).

The procedure is then to use a matrix element generator to generate
exclusive $n$-parton final states, to reweight these states with the
Sudakov form factors, and then add a parton cascade.

Note, however, that the $\rho$ scales are defined in a particular
parton shower scenario and have no equivalents in the matrix element
generation, where the notion of one emission being performed before
another is not well defined. Instead these scales need to be
reconstructed from the generated $n$-parton state.

In reference \cite{Catani:2001cc} this is achieved by using the
\kT-algorithm\cite{kTalgorithm,Catani:1991hj}, the Sudakov form
factors are then calculated according to an analytical form, and the
angular-ordered parton shower in \herwig\cite{Corcella:2000bw} is
added with a special veto procedure. The reason for this veto
procedure is to avoid double counting, since the reconstructed scales
do not correspond to the ordering variable in the parton shower. The
angular ordering means that one can have a soft emission at a large
angle, followed by a harder emission at a smaller angle.

Here the Colour Dipole Model implemented in the \ariadne\ program will
be used for the parton cascade. The ordering variable is an invariant
transverse momentum, so the emissions are ordered in scale.  In
addition, a modified version of the
\diclus\cite{Lonnblad:1993qd,Moretti:1998qx} jet clustering algorithm
will be used to reconstruct the emission scales given as the same
invariant transverse momenta. In this way many of the problems in
ref.~\cite{Catani:2001cc} are completely avoided. The \diclus\ 
algorithm will not only be used to reconstruct the emission scales,
but to reconstruct a complete dipole cascade \textit{history} for the
$n$-parton states generated according to the matrix element, i.e.\ 
also the intermediate states which would occur in a dipole shower
producing the same state.

These states will be used to evaluate the Sudakov form factors in
exactly the same way as they are evaluated in the dipole cascade as
explained in section \ref{sec:sudak-veto-algor}.  In the procedure of
Catani et al.\ the Sudakovs are calculated analytically, which means
that they do not exactly correspond to a no-emission probability since
they do not take into account that the available phase space for
emission off a quark after it has emitted a gluon is changed due to
recoils. In the dipole cascade the Sudakovs are generated dynamically
and takes into account the actual phase space available for further
emissions. To MLLA accuracy the two approaches are, however,
equivalent.

\section{Reconstruction of emissions}
\label{sec:reconstr-emiss}

The colour-dipole cascade
model\cite{Gustafson:1986db,Gustafson:1988rq} describes the emission
of a gluon in terms of dipole radiation from a colour dipole between
two partons.  The emissions are hence described as two partons going
to three, rather than one going to two as in conventional parton
shower models.  This means that gluon coherence is automatically taken
into account and that the first gluon emission in \ee\ annihilation
trivially reproduces the full first order matrix element. But there
are a couple of technical details which are particular to the dipole
cascade. All partons are always on-shell at each step of the cascade.
The conservation of energy and momentum is achieved since both
emitting partons receives a recoil from the emitted gluon. The
splitting of a gluon into a \q\qbar-pair is also treated as if emitted
from one of the dipoles connected to the gluon, and the parton in the
other end will receive some recoil in order to conserve anergy and
momentum\cite{CDMsplit90}.  Further more, the scale of an emission is
defined in terms of a Lorentz-invariant \pT\ of the emitted parton
with respect to the emitting ones. This is defined as
\begin{equation}
  \label{eq:invpt}
  \pT^2=\frac{(s_{12}-(m_1+m_2)^2)(s_{23}-(m_2+m_3)^2)}
  {s_{123}},
\end{equation}
where parton 2 is the emitted one and $s_{ij}$ and $s_{ijk}$ are the
squared invariant masses of the two- and three-parton combinations.

The dipole clustering algorithm
(\diclus)\cite{Lonnblad:1993qd,Moretti:1998qx} can be thought of as
the inverse of the dipole cascade. In each step the combination of
three jets which have the smallest invariant \pT\ are clustered
together into two (massless) jets. In the procedure presented here,
the dipole clustering will be used with a couple of modifications.
First of all the information available from the generation of the
few-parton state will be used. Hence a gluon is only considered to
have been emitted from the two partons to which it is colour
connected\footnote{This requires that the ME generator produces states
  with a definite colour topology. Although this information is not
  physical, it is usually provided by ME generator programs. See
  discussion in ref.~\cite{Boos:2001cv}}. So for a colour connected
$\q\g_1\g_2\qbar$ state the only possible clusterings are
$\q\g_1\g_2\to\q\g_2$ and $\g_1\g_2\qbar\to\g_1\qbar$. Also, a
\q\qbar-pair is reconstructed into a gluon which is made massless by
also considering one of the partons connected to the \q\ or \qbar. In
this way all partons in the reconstructed states are always kept
on-shell. A \q\qbar\ pair is only reconstructed to a gluon if they are
not colour-connected (directly or indirectly).

When reconstructing a 2\to3 emission there is an ambiguity in the
directions of the emitting partons. If the three partons are
transformed into their center of mass system, the momenta of the
emitting partons are easily obtained from energy momentum
conservation. It is also clear that the two partons should lie in the
same plane as the original three. But the angular orientation in this
plane is not determined. The inverse problem is encountered in the
dipole cascade where the amount of transverse recoil taken by each of
the emitting partons is not given by the theory. There are different
choices made for each kind of emission. In \ariadne\ the choices are
as follows:
\begin{itemize}
  \itemsep -2mm
\item For a gluon emission from a \q\qbar\ dipole, one of the quarks
  retains its direction with a probability proportional to the square
  of its energy (according to the prescription in
  \cite{Kleiss:1986re}).
\item For gluon emission from a quark-gluon dipole, the gluon retains
  its direction.
\item For gluon emission from a gluon-gluon dipole, the transverse
  recoil is shared among the emitters so that the sum of their squared
  transverse momenta is minimized.
\item For a gluon splitting into a \q\qbar\ pair, the spectator parton
  retains its direction.
\end{itemize}

Of these, all but the first are completely deterministic and can
easily be inverted and be used in the reconstruction algorithm. For
the first, the prescription for gluon-gluon dipoles is followed
instead\footnote{This will not influence the results in this paper
  where only $\as^2$ ME will be used and all $\q\g\qbar\to\q\qbar$
  reconstructions will give the initial \q\qbar\ state where the
  angular orientation is irrelevant for the subsequent cascade.}.

Rather than always selecting the three parton configuration which has
the smallest invariant \pT\ to be reconstructed, as is customary in
jet algorithms, it is possible to reconstruct all possible dipole
cascade histories. This is feasible since we are dealing with only a
handful partons. The procedure will then be to choose randomly between
these different histories weighted with the corresponding dipole
splitting probabilities in analogy to the strategy in
\cite{Andre:1998vh}. The splitting probabilities will not include a
running \as\ as in the normal dipole cascade, since a constant \as\ 
was used in the generation of the parton state. The running of \as\ 
will be corrected for at a later stage.

It should be noted that some of the histories may consist of sequences
of un-ordered emissions which cannot have been produced by the dipole
cascade, and these are excluded from the histories to choose between.
In rare cases it is possible that no history can be found which
correspond to an ordered sequence of dipole emissions. In this case
one of the ``impossible'' histories are chosen, but the reconstructed
scales are modified so that if the scale of one emission is smaller
than the scale of the subsequent one, the larger scale is chosen for
both reconstructed emissions.

With this procedure it is now possible to reconstruct a dipole cascade
history for any $n$-parton state. All intermediate $2$, $3$, \ldots,
$n-1$ parton states are reconstructed together with the corresponding
emission scales $\pTi{1}^2$, \ldots, $\pTi{n-2}^2$. The reconstructed
scales can then be used to correct the MEs for the running of \as\ by
rejecting the state with a probability
\begin{equation}
  \label{eq:asreject}
  1-\frac{1}{\asi{0}^{n-2}}\Pi_{i=1}^{n-2}\as(\pTi{i}^2)  
\end{equation}
where the \asi{0} used in the ME generation is taken at the cutoff
scale $\pTi{c}^2$\ of the parton cascade\footnote{Note that the $Q_0$
  cut used in the ME generation is not necessarily the smallest
  possible invariant \pT\ scale.} to ensure that the probability
is positive.

\section{The Sudakov Veto algorithm}
\label{sec:sudak-veto-algor}

The reconstructed scales and states are also used to calculate the
correction for the Sudakov form factors. Rather than using the
approximate analytic expression as a weight, we can use the fact
that it corresponds to the no-emission probability in a specific
region of phase space.

Consider a three-parton state generated with the \ordo{\as} ME, where
the scale of the gluon emission has been reconstructed to $\pTi{1}^2$.
The Sudakov form factor is then the probability of there being no
emission from the initial \q\qbar\ state \emph{before} the gluon was
emitted, i.e.\ at a scale above $\pTi{1}^2$, and that there is no
emission from the \q\g\qbar\ state between the scale $\pTi{1}^2$ and
the cutoff in the ME. By making two trial emissions with the dipole
cascade, one from the reconstructed \q\qbar\ state, starting from the
maximum scale, and one from the ME-generated \q\g\qbar\ state starting
from $\pTi{1}^2$ and rejecting \emph{the whole event} if the first was
at a scale above $\pTi{1}^2$ or the second was inside the ME cutoff,
the probability of accepting the event is exactly equal to the Sudakov
form factor. With this veto procedure the proper phase space region is
taken into account rather than the approximate limits in the analytic
form.

Special care must also be taken when a partonic state has been
generated to the highest order in \as\ used for the ME generation. In
that case we want to continue generating with the dipole cascade from
the last reconstructed scale irrespectively of the cutoff in the MEs
and we should keep the trial emission from the ME generated state and
continue the cascade rather than vetoing the whole event. This
corresponds to changing the four-jet rates in (\ref{eq:RnMEPS}) to
four-or-more-jets rates, e.g.
\begin{eqnarray}
  \label{eq:r4plus}
  R_{4gg+}(\rho_m,\rho_0) &=&
  \as^2\int^{\rho_m}_{\rho_0}d\rho_1\int^{\rho_1}_{\rho_0}d\rho_2\,\,
  c\sup{ME}_{2,2(gg)}(\rho_m,\rho_1,\rho_2)\times\\
  & &\rule{38mm}{0mm}\sud{2}(\rho_{m},\rho_1)
  \sud{3}(\rho_1,\rho_2).\nonumber
\end{eqnarray}
This cannot easily be achieved in the procedure of Catani et al.,
since the angular ordering of the parton shower in \herwig, allows for
subsequent harder emissions which could result in double counting.

\section{The algorithm step by step}
\label{sec:procedure-step-step}

We now have all the ingredients to present the whole algorithm step by
step. We assume there is a matrix element generator which can generate
complete partonic states according to the exact tree-level MEs up to
\ordo{\as^N}. This generator is regulated by a cutoff $Q_0$ to avoid
divergencies. In principle this cutoff could be in the same invariant
\pT\ used as evolution variable in the dipole cascade, but to be
completely general we assume that it is instead e.g.\ a simple cutoff
in invariant mass of any two outgoing partons. This is allowed as long
as there is no double counting (or \textit{under counting} due to
empty phase space regions) as is demonstrated below. This matrix
element generator should then be combined with the standard dipole
cascade of \ariadne\ which has a lower cutoff in the invariant \pT\ 
given by $\pTi{c}$. The constant \asi{0} used in the ME generator is
taken at the scale $\pTi{c}^2$ using the same \lam\ as in the dipole
cascade.

The whole procedure will now be as follows:
\begin{enumerate}\itemsep -1.5mm
\item First the number of partons, $n\le N$, to be generated is chosen
  according to the integrated jet rates $R_n$ from the tree-level MEs.
  Note that $\sum R_n$ is larger than $1$ since we do not include
  virtual corrections.
\item Then the momenta of the $n$ partons are generated according to
  the \ordo{\as^{n-2}} tree-level ME. Afterwards the invariant $\pT^2$ of
  the $n$ partons is checked, and if anyone is below $\pTi{c}^2$, the
  state is rejected and the procedure is restarted at step 1.
\item Now, all the intermediate states $S_2$, \ldots, $S_{n-1}$ and
  scales $\pTi{1}^2$, \ldots, $\pTi{n-2}^2$ corresponding to a
  sequence of dipole emissions are reconstructed according to the
  algorithm in section \ref{sec:reconstr-emiss}.
\item The generated event is rejected and we restart at step 1 with a
  probability given by equation (\ref{eq:asreject}).
\item We now make a trial emission with the dipole cascade from the
  state $S_2$ starting from the maximum scale limited by the center of
  mass energy. If this emission is at a scale above $\pTi{1}^2$, the
  event is rejected and we restart from step 1. If not, a trial
  emission is performed from the state $S_3$ with a maximum scale of
  $\pTi{1}^2$.  If this emission is at a scale above $\pTi{2}^2$ the
  event is rejected and we restart from step 1. This procedure is
  repeated for all states down to $S_{n-1}$.  If no rejection has been
  made, a trial emission is made from the ME-generated $n$-parton
  state starting from the scale $\pTi{n-2}^2$.  There are now two
  cases
  \begin{itemize}\itemsep 0mm
  \item If $n=N$ the trial emission is always kept and the dipole
    cascade is allowed to continue down to the cutoff $\pTi{c}^2$ and
    the event is accepted.
  \item If $n<N$, and all parton pairs pass the ME cut, $Q_0$, the
    event is rejected and we restart from step 1. If any of the
    partons fail the cut, the trial emission is accepted and the
    dipole cascade is allowed to continue down to the cutoff
    $\pTi{c}^2$ and the event is accepted.
  \end{itemize}
\end{enumerate}

\begin{figure}[t]
  \centering
  \input{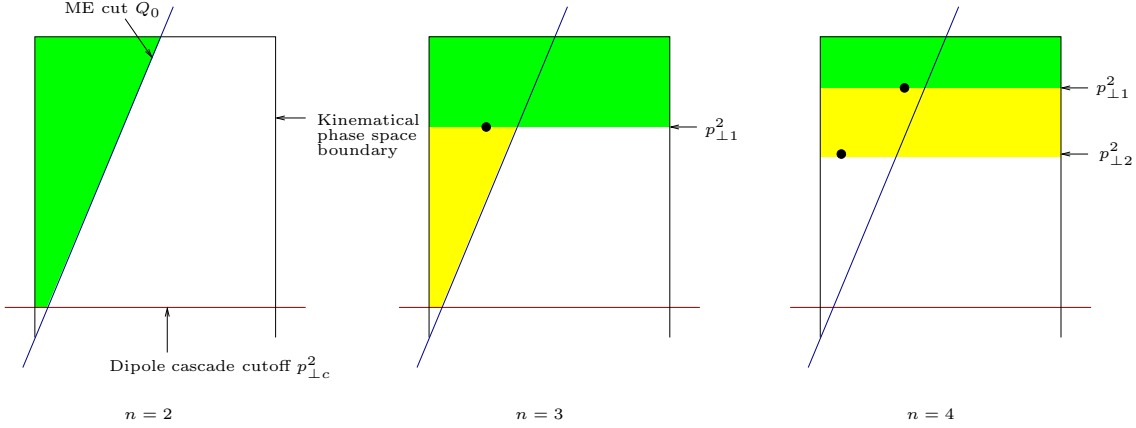}
  \caption[dummy]{{\it The integration regions (shaded) for the Sudakov
      form factors in a symbolic two-dimensional phase space, giving
      the probability to accept a ME-generated 2, 3 and 4 parton
      states in the case $N=4$. The evolution variable in the cascade
      is assumed to be along the vertical axis, while the horizontal
      axis is some rapidity or energy splitting variable. The cutoff
      in the ME is along the diagonal line.}}
  \label{fig:phase}
\end{figure}

\begin{figure}
  \centering
  \input{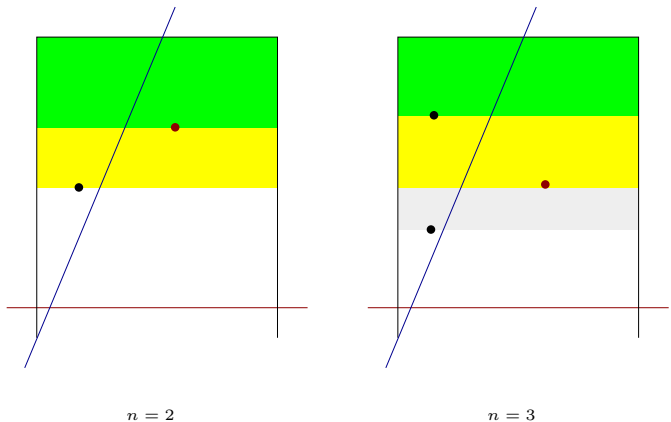}
  \caption[dummy]{{\it Possible sequences of dipole cascade emissions
      from $n=2$ and $n=3$ states generated with MEs.}}
  \label{fig:phase2}
\end{figure}

For a pictorial description of the procedure, figure \ref{fig:phase}
shows the regions in a symbolic two-dimensional phase space, in which
the Sudakov veto algorithm is used in the case of $N=4$ and $n=2,3,4$.
It is clear that if there is only one emission inside the ME cut, it
is handled by the ME generator and this region is never populated by
the dipole cascade. If there are two emissions inside the ME cut, the
two hardest ones are always handled by the ME generator while
additional emissions are given by the dipole cascade. In this way
there is no double counting between the ME generator and the dipole
cascade. Note that if the trial emission in e.g\ the $n=2$ case is
outside the ME cut and therefore is accepted as shown in figure
\ref{fig:phase2}, further emissions inside the ME cut are allowed
without any risk of double counting.

Although the procedure is to add a dipole cascade to an $N$-parton
matrix element generator, the result is that all final multi-parton
states are distrubuted as if generated by the dipole cascade, except
that if the $n-2$ (with $n\leq N$) hardest emissions are inside the
matrix element cut, their distribution is described by the exact
tree-level matrix element.

More precisely, we have the following possible situations:
\begin{itemize}
\item If the hardest emission is outside the ME cutoff, all emissions
  are given by the dipole splitting functions, even if softer
  emissions are inside.
\item If the hardest emission is inside the ME cutoff, but the second
  hardest is outside, the hardest emission will be given by the ME,
  while all subsequent ones are given by the dipole splitting
  functions.
\item If the two hardest emissions are inside the ME cutoff they are
  given by the ME, while all subsequent emissions are given by the
  dipole splitting functions (irrespective of whether they are inside
  the cutoff or not).
\end{itemize}
In all cases the Sudakov form factors are the same irrespective of
whether the emissions are given by the ME or the dipole splitting
functions.

It is clear that there should only be a small dependence on the $Q_0$
since the only change when going outside the cut is that the emissions
are governed by the leading logarithmic expressions rather than the
exact ME and these should be very similar for a small enough cut. It
is important that the extra \pT\ cut used in step 2 is the same for
the ME-generated state as for the subsequent dipole cascade, otherwise
the ME may populate phase space outside of the reach of the cascade.
This is, in fact, not quite trivially achieved in the cascade itself.
The reason is that the recoils from an emission in one dipole may push
the invariant \pT\ of a parton in a neighboring dipole below the
cutoff \pTi{c}.  To be completely consistent, the dipole cascade is
therefore used with an extra cut vetoing emissions which push other
partons below the cut.

\section{Results}
\label{sec:results}

To check that the procedure works, the simplest thing is to test it
for $N=3$ since the \ariadne\ dipole cascade is already corrected to
match the \ordo{\as} ME. And, indeed the new procedure completely
agrees with standard \ariadne\ in this case. A less trivial test is to
look at the $N=4$ case. For this a modification of the \ordo{\as^2} ME
generator implemented in \pythia\ is used, stripped down so that it
only uses the tree-level MEs with a simple invariant mass cutoff. This
is then used together with the new dipole cascade interface algorithm
implemented in \ariadne.

\begin{figure}[t]
  \centering

  \hspace*{-8mm}\begin{minipage}{16cm}
    \epsfig{figure=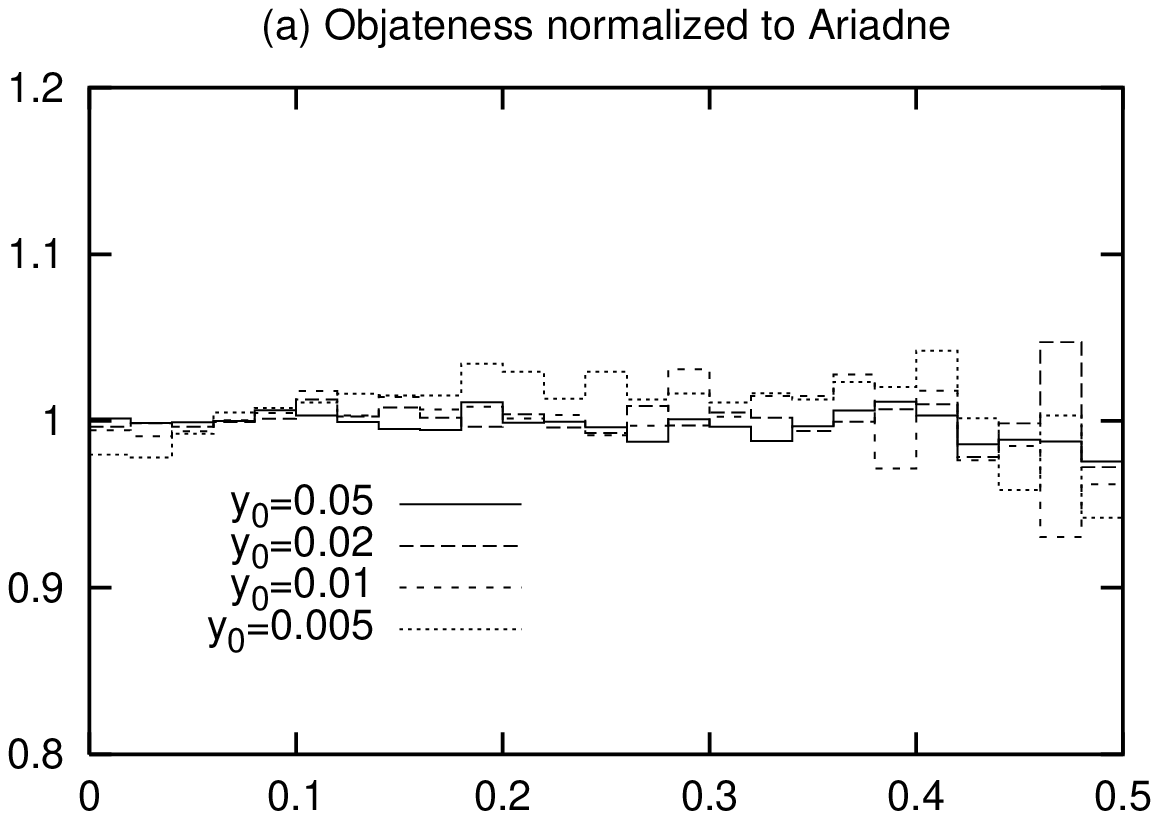,width=5.5cm,angle=-90}
    \epsfig{figure=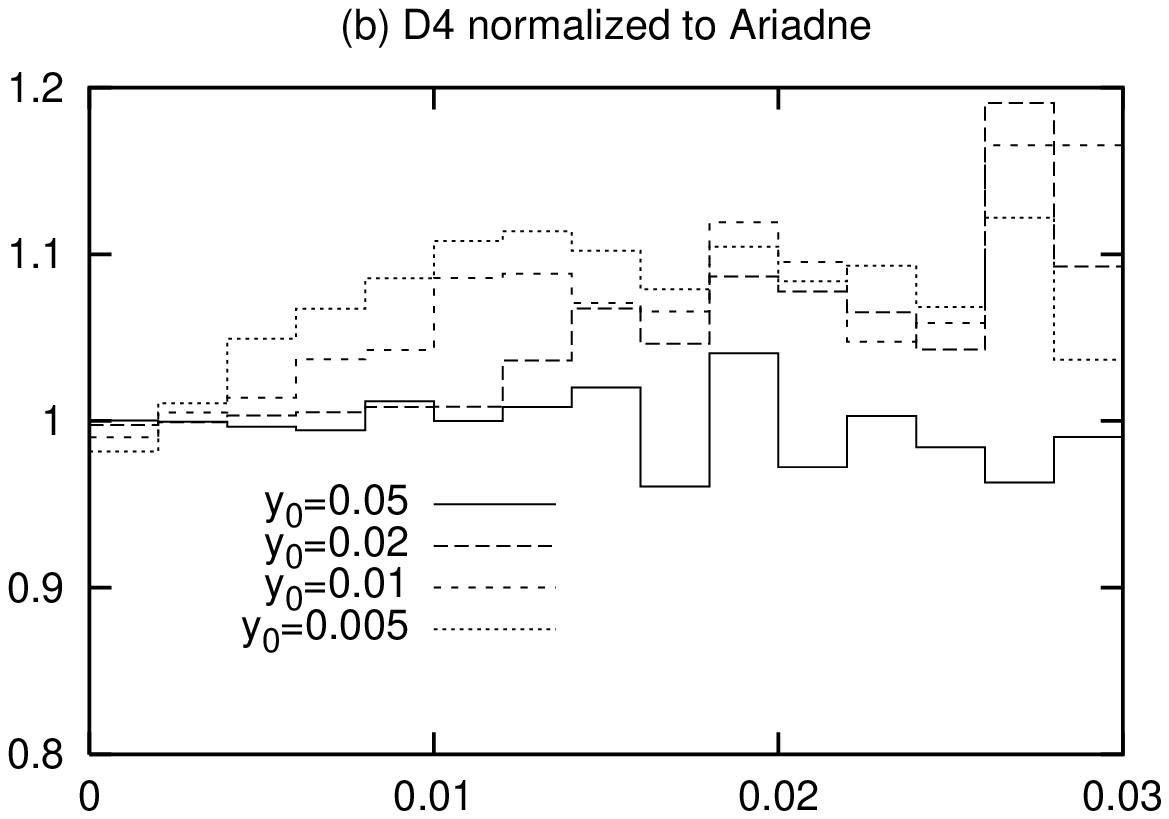,width=5.5cm,angle=-90}

    \epsfig{figure=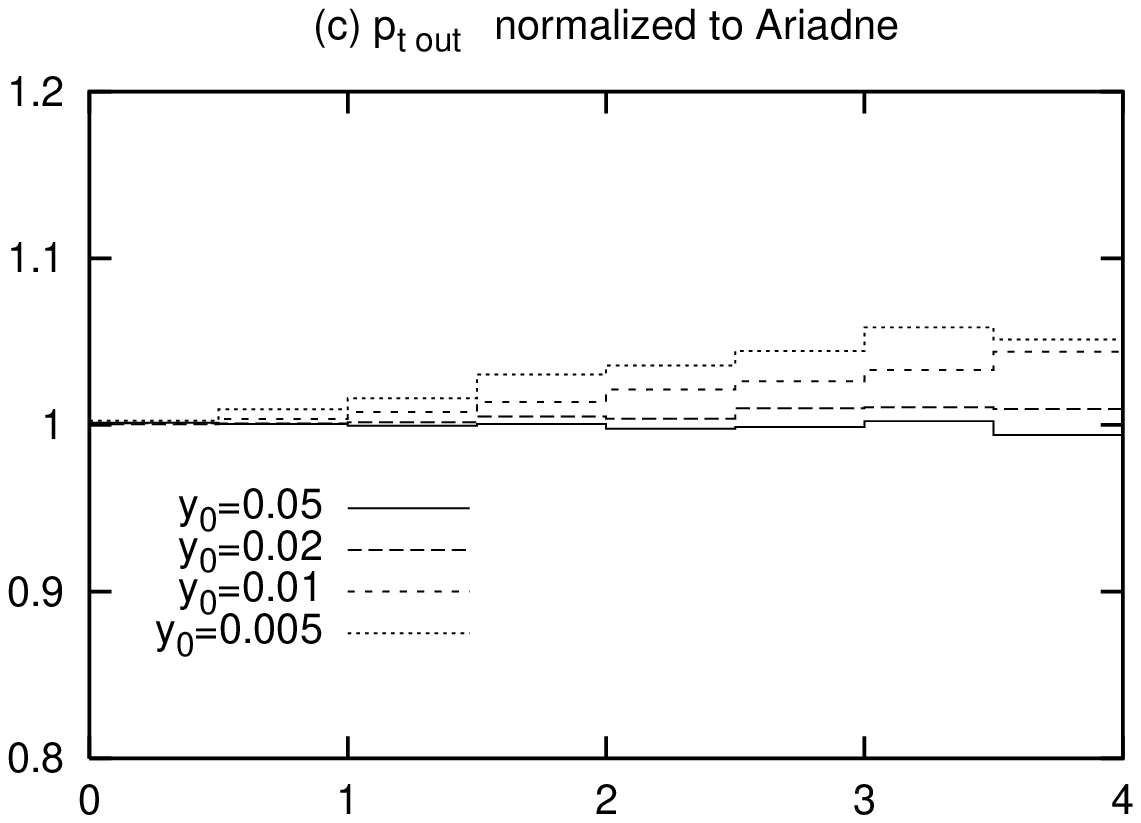,width=5.5cm,angle=-90}
    \epsfig{figure=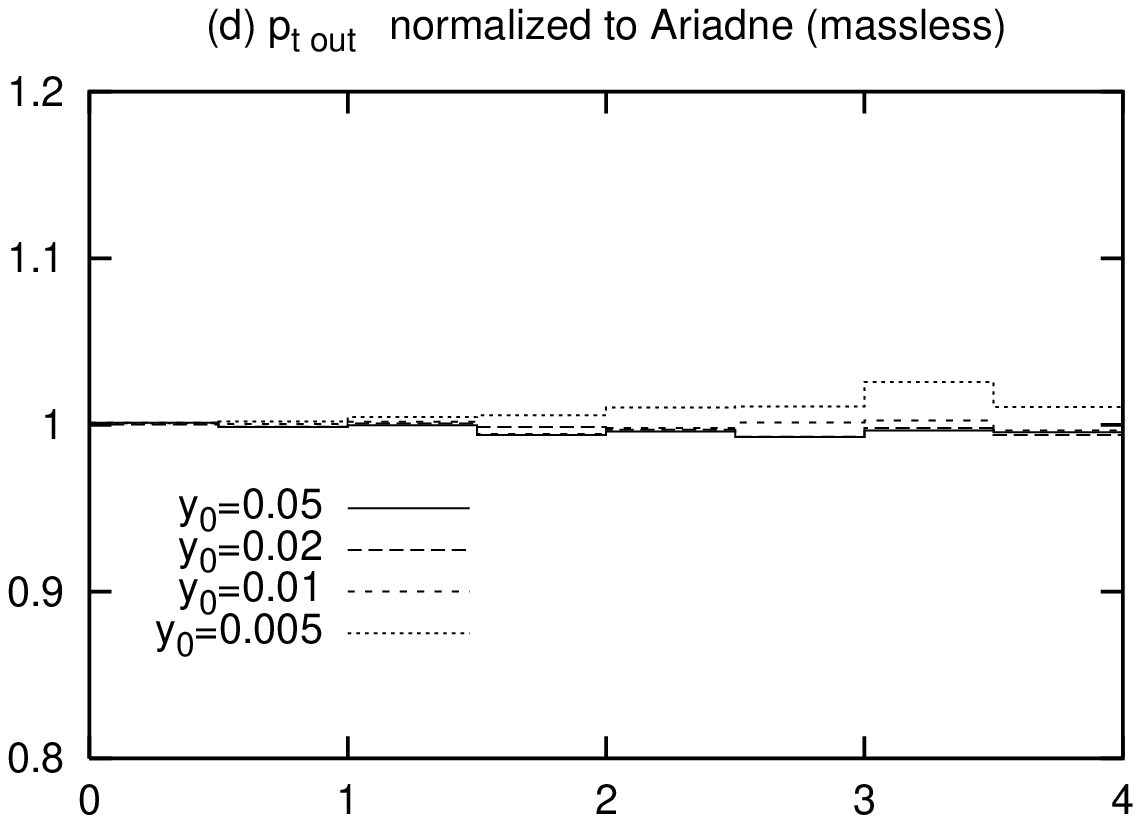,width=5.5cm,angle=-90}
  \end{minipage}

  \caption[dummy]{{\it Ratios of parton level event shapes at
      $E\sub{CM}=91$~GeV for the new ME matching algorithm using
      different values of $Q_0$ w.r.t.\ to the standard \ariadne\ 
      program. The distributions are (a) oblateness, (b) 4-jet
      resolution value for the Durham jet algorithm and (c)
      \pTi{\mbox{out}}. (d) is the same as (c) but with massless
      quarks and no secondary quarks both for the new ME matching and
      for standard \ariadne. In all cases, the full line is with
      $y_0=Q^2_0/Q^2=0.05$, long-dashed: $y_0=0.02$, dashed:
      $y_0=0.01$ and dotted: $y_0=0.005$. }}
  \label{fig:shapes}
\end{figure}

First we look at some standard event shapes.  These are known to be
very well described by the \ariadne\ program (see eg.\ 
\cite{Hamacher:1995df}). We will not directly compare to data since
this would not be meaningful without a retuning of the hadronization
parameters.  Instead we will compare the new procedure with the
standard \ariadne\ program on parton level.  Figure \ref{fig:shapes}
shows the dependence on the ME cut $Q_0$ of some standard event shapes
which should be sensitive to \ordo{\as^2} effects. The differences
w.r.t.\ \ariadne\ are very small and rather than showing the event
shapes them selves, figure \ref{fig:shapes} shows the ratio to the
\ariadne\ results. The fact that the differences are so small is not a
surprise.  Already in \cite{Andersson:1992he} it was shown that the
dipole cascade agrees very well with \ordo{\as^2} MEs in most regions
of phase space.  For large $Q_0$ the new procedure reduces to the
standard \ariadne\ program as expected.  The dependence on the $Q_0$
is small but not zero.  The largest difference is found i the D4
distribution which measures the smallest distance, according to the
Durham jet-algorithm \cite{kTalgorithm,Catani:1991hj}, between two
jets when an event has been clustered to four jets. It is a bit
worrying that the dependence does not seem to disappear even for very
small $y_0=Q_0^2/E\sub{cm}^2$.  This may be expected for the D4
distribution when the distance between jets are smaller than $y_0$,
but it is also present in the \pTi{\mbox{\scriptsize out}}
distribution even for fairly large values. The reason can be traced to
the treatment of quark masses which is not exact in \ariadne\ (nor is
it exact in the tree-level matrix elements in \pythia). In addition
there are some uncertainties in the treatment of secondary quarks in
\ariadne\footnote{An investigation of the effects of quark masses and
  secondary quarks will be presented in a future
  publication\cite{Carlosprep}}. As seen in in figure
\ref{fig:shapes}d, if all quark masses are put to zero and secondary
quarks are taken away, the dependency on $y_0$ becomes much smaller,
but it still does not go away for really small values. A possible
explanation may be that the dipole clustering routine has some
problems for very small resolution scales as described in
\cite{Moretti:1998qx}, but the dependence is in any case much smaller
than the uncertainties due to hadronization parameters and the basic
parameters in the dipole cascade, \pTi{c} and \lam.

\begin{figure}
  \centering
  \hspace*{-8mm}\begin{minipage}{16cm}
    \epsfig{figure=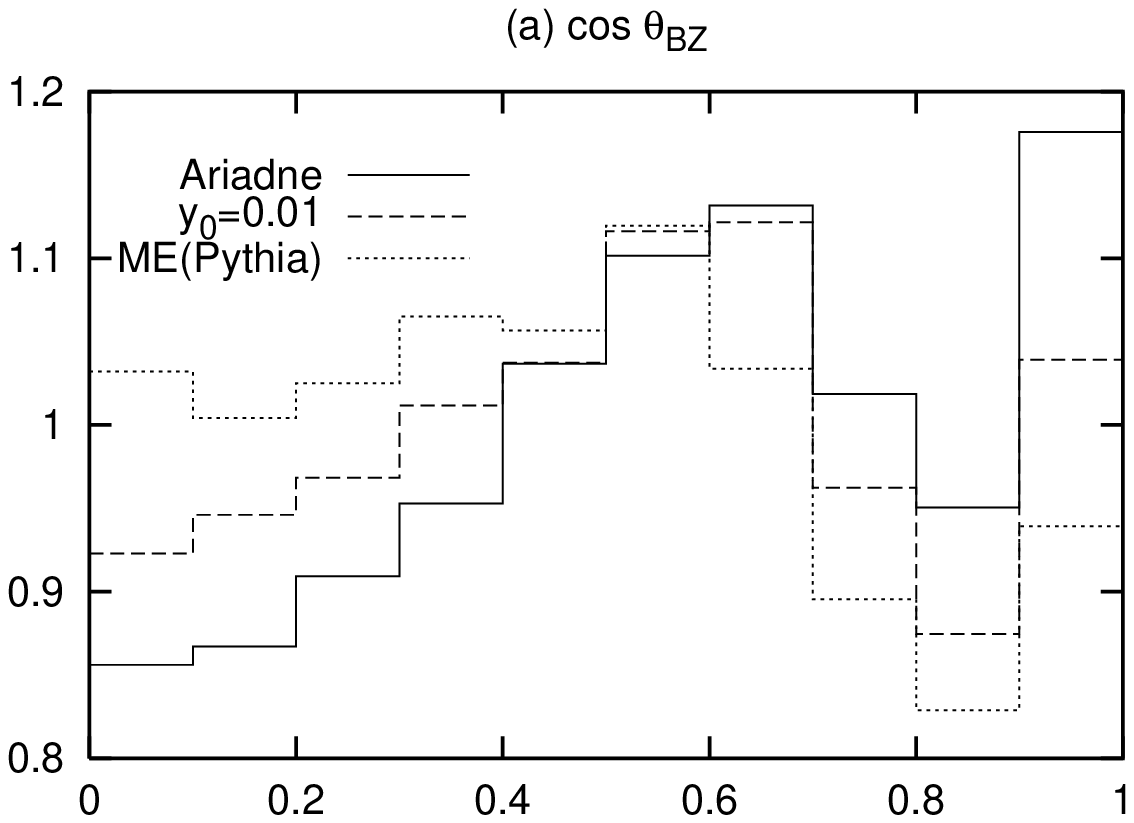,width=5.5cm,angle=-90}
    \epsfig{figure=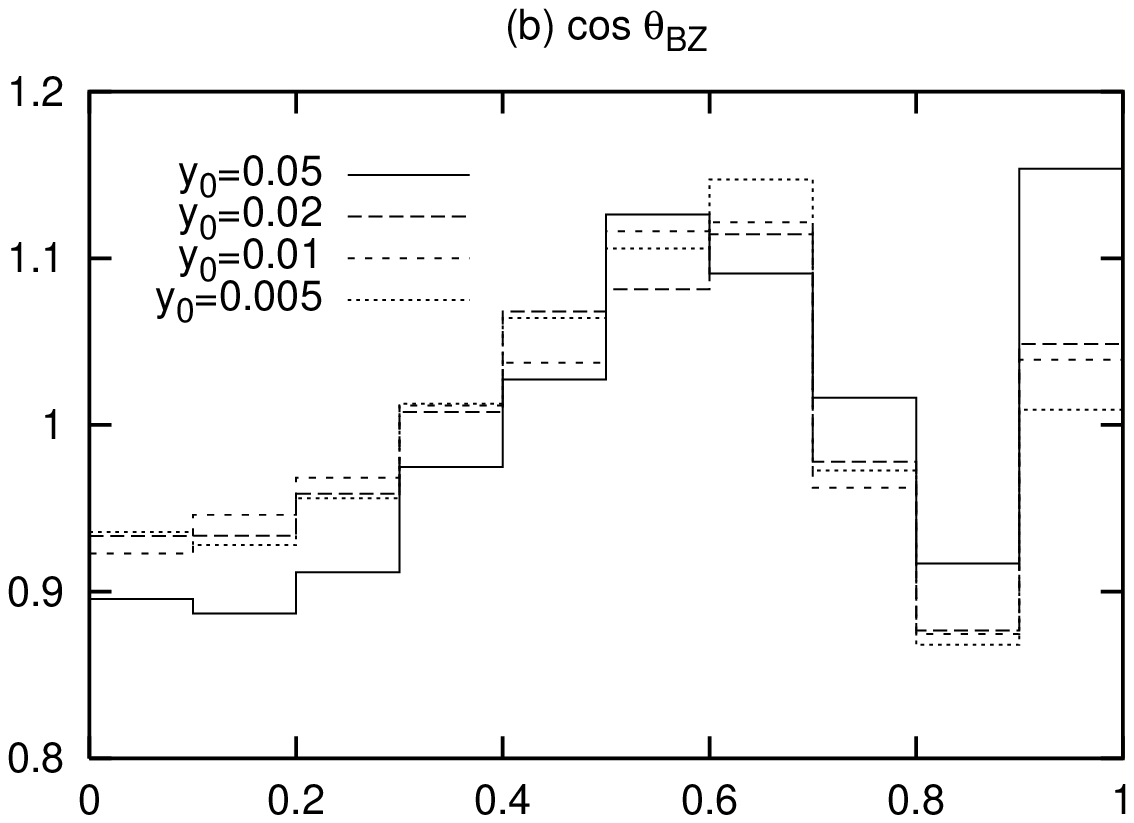,width=5.5cm,angle=-90}
  \end{minipage}
  \caption[dummy]{{\it The distribution in the Bengtsson--Zerwas
      angle on parton level using the JADE jet
      algorithm\cite{Bartel:1987mf} with $y\sub{cut}=0.03$.  In (a)
      the full line is standard \ariadne, the dashed line is the new
      ME matching algorithm with $y_0=0.01$ and the dotted line is the
      tree-level \ordo{\as^2} ME-only generator in \pythia\ with
      $y_0=0.01$. In (b) the full, long dashed, dashed and dotted
      lines are all the new ME matching algorithm with
      $y_0=0.05,0.02,0.01$ and $0.005$ respectively.}}
  \label{fig:BZ}
\end{figure}

To really see the influence of the ME matching one must look at
details in the correlations between jets. Here we will look at
Bengtsson--Zerwas angle \cite{Bengtsson:1988qg}, i.e.\ the angle
between the plane spanned by the two largest jets and the plane
spanned by the two smallest jets in a four-jet event, which is not at
all described by the standard \ariadne\ program. In figure
\ref{fig:BZ} we see that the new matching procedure is closer than
standard \ariadne\ to the result from the pure \ordo{\as^2} ME
generator in \pythia. It does not, and should not, exactly reproduce
the pure ME approach since the correlation is smeared by the
subsequent soft radiation.

\section{Conclusions and outlook}
\label{sec:conclusions-outlook}

The procedure presented here to correct the dipole cascade in
\ariadne\ to match the exact \ordo{\as^N} matrix elements works. The
only additional parameter needed is the cutoff $Q_0$ in the ME
generation, but the results have been shown to be fairly insensitive
to the value chosen as long as it is reasonably small. Although the
results presented here has only been for $N=4$, the procedure is
completely general and can be applied to basically any $N$-parton ME
generator. One advantage is that the procedure is practically
non-intrusive with respect to the ME generator used. On the other hand
the vetoing technique makes the generation very inefficient for very
small $Q_0$.

There are a few uncertainties in this algorithm, mainly to do with the
reconstructions of the emissions. It is not quite clear what to do
with reconstructed emissions which are not ordered. The procedure
presented in section \ref{sec:reconstr-emiss} is not unique and, in
fact, several other options have been tested. The results are,
however, rather stable w.r.t.\ such variations.

In any case the procedure will give multi parton final states where
the first \makebox{$N-2$} emissions are correct to exact
\ordo{\as^{N-2}} accuracy and all others are correct to MLLA accuracy.

It could be possible to also include the exact virtual corrections to
\ordo{\as^{N-2}}. This would involve correcting the tree-level matrix
elements with the difference between the exact and the leading
logarithmic parts of the virtual plus unresolved terms, for the $N=4$
case effectively turning $R_2$ and $R_3$ in equation (\ref{eq:RnMEPS})
into
\begin{eqnarray}
  \label{eq:RnMEPSExact}
  R_2(\rho_m,\rho_0) &=&\sud{2}(\rho_{m},\rho_0) +
  \as\delta C\sup{MEPS}_{0,1}(\rho_m,\rho_0) +
  \as^2\delta C\sup{MEPS}_{0,2}(\rho_m,\rho_0),\nonumber\\
  R_3(\rho_m,\rho_0) &=& \as\int^{\rho_m}_{\rho_0}d\rho\,\,
  c\sup{ME}_{1,1}(\rho_m,\rho)\times\\
  & & \left[\sud{2}(\rho_{m},\rho)\sud{3}(\rho,\rho_0) +
  \as\delta  C\sup{MEPS}_{1,2}(\rho_m,\rho,\rho_0)\right],\nonumber\\
\end{eqnarray}
where the difference between the full coefficients from the summed
virtual and unresolved diagrams in the matrix element and the
corresponding parton shower ones, from the expansion of the Sudakov
form factors and the running of \as\ with transverse momenta in the
parton cascade, are symbolically written $\delta c\sup{MEPS}_{n,m} =
c\sup{ME}_{n,m} - c\sup{PS}_{n,m}$.  These coefficients are free from
singularities, but the cutoff in the matrix element generation must be
chosen so that the differential $n$-jet rates will come out positive
in the whole phase space. Further investigations is needed to see if
this at all can be accomplished.

It should be noted that a procedure for combining next-to-leading
order matrix element with parton showers has recently been proposed by
Frixione and Webber \cite{Frixione:2002ik}. However, such a generator
will produce negatively weighted events and is therefore difficult to
use as a general purpose event generator.

Finally it should be noted that it is possible to use the procedure
presented in this paper also for collisions with incoming hadrons. The
additional complication is that the reconstruction of the intermediate
states and scales must take into account that there may both be
initial- and final-state emissions, where the former depends on the
parton density distributions of the incoming hadrons.  But otherwise
the procedure would be the same: Use a tree-level matrix element
generator to generate lowest order and subsequent higher order
partonic states, reconstruct the possible emission histories according
to a parton cascade scheme together with the corresponding emission
scales, calculate the running of \as, calculate the Sudakov form
factors with the veto algorithm using trial emissions from the
reconstructed states, and continue the cascade.

\bibliographystyle{utcaps}  
\bibliography{references} 

\end{document}